\documentclass[prl,twocolumn,floatfix,amsmath,amssymb,superscriptaddress]{revtex4}
\usepackage{graphicx,color}
\usepackage{dcolumn}
\usepackage{bm}
\begin{document}

\title{Low temperature broken symmetry phases of spiral antiferromagnets}
\author{Luca Capriotti}
\affiliation{Credit Suisse First Boston (Europe) Ltd.,
One Cabot Square, London E14 4QJ, United Kingdom}
\affiliation{Kavli Institute for Theoretical Physics, University
of California, Santa Barbara, California 93106-4030}

\author{Subir Sachdev}
\affiliation{Department of Physics, Yale University, P.O. Box
208120, New Haven, CT 06520-8120}
\date{September 19, 2004}

\begin{abstract}
We study Heisenberg antiferromagnets with nearest- ($J_1$) and
third- ($J_3$) neighbor exchange on the square lattice. In the
limit of spin $S \rightarrow \infty$, there is a zero temperature
($T$) Lifshitz point at $J_3 = \frac{1}{4} J_{1}$, with long-range
spiral spin order at $T=0$ for $J_3
> \frac{1}{4} J_{1}$. We present classical Monte Carlo simulations and
a theory for $T>0$ crossovers near the Lifshitz point: spin
rotation symmetry is restored at any $T>0$, but there is a broken
lattice reflection symmetry for $0 \leq T < T_c \sim (J_3 -
\frac{1}{4} J_1)S^2$. The transition at $T=T_c$ is consistent with
Ising universality. We also discuss the quantum phase diagram for
finite $S$.
\end{abstract}
\pacs{pacs here}

\vspace{3cm}




\maketitle

Frustrated antiferromagnets have recently attracted much interest
in connection with the possibility of stabilizing unconventional
low temperature ($T$) phases, possibly falling outside the known
paradigms of condensed matter physics \cite{dcq}. In this respect,
the most interesting systems are those where the combined effect
of frustration and quantum fluctuations is strong enough to
prevent the onset of magnetic ordering, thus stabilizing `exotic'
quantum disordered ground states \cite{sl}. A very promising
candidate for such a {\sl spin-liquid} phase has been found in the
two-dimensional $J_1{-}J_3$ model defined by the following
Hamiltonian
\begin{equation} \label{ham}
\hat{H}=J{_1}\sum_{\langle i,j\rangle}
\hat{{\bf {S}}}_{i} \cdot \hat{{\bf {S}}}_{j}
+ J{_3}\sum_{\langle\langle i,j\rangle\rangle}
\hat{{\bf {S}}}_{i} \cdot \hat{{\bf {S}}}_{j}~~,
\end{equation}
where,  $\hat{{\bf {S}}}_{i}$ are spin-$S$ operators on a square
lattice and $J_1,J_3\ge 0$ are the nearest- and third- neighbor
antiferromagnetic couplings along the two coordinate axes. For
this model, early large $N$ computations, \cite{rs} and recent
large scale Density Matrix Renormalization Group (DMRG)
calculations for $S=1/2$ \cite{j1j3} have suggested the existence
of a gapped spin liquid state with exponentially decaying spin
correlations and no broken translation symmetry in the regime of
strong frustration ($J_3/J_1\simeq 0.5$).

\begin{figure}
\vspace{-0mm}
\includegraphics[width=0.47\textwidth]{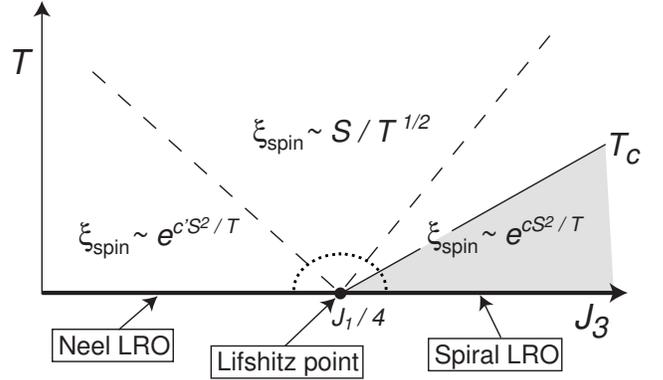}
\vspace{0mm} \caption{\label{phasediag} Phase diagram of $\hat{H}$
in the limit $S\rightarrow \infty$. The shaded region has a broken
symmetry of lattice reflections about the $x$ and $y$ axes,
leading to Ising nematic order. The Ising transition is at the
temperature $T_c \sim (J_3 - \frac{1}{4} J_1) S^2$. The spin
correlation length, $\xi_{\rm spin}$, is finite for all $T>0$,
with the $T$ dependencies as shown, with $c/2 = c' = 8\pi |J_3 -
\frac{1}{4} J_1|$; the crossovers between the different behaviors
of $\xi_{\rm spin}$ are at the dashed lines at $T \sim |J_3 -
\frac{1}{4} J_1| S^2$. Spin rotation symmetry is broken only at
$T=0$ where $\xi_{\rm spin}=\infty$. There is no Lifshitz point at
finite $S$ because it is pre-empted \cite{il} by quantum effects
within the dotted semi-circle: here there is a $T=0$ spin gap
$\Delta \sim S \exp(-\tilde{c} S)$ and spin rotation symmetry is
preserved. This semicircular region extends over $T \sim |J_3 -
\frac{1}{4} J_1| S \sim \Delta$. Further details on the physics
within this region appear at the end of the paper.}
\end{figure}

This paper will describe properties of the above model for large
$S$, and discuss consequences for general $S$. Our results,
obtained by classical Monte Carlo simulations and a theory
described below, are summarized in Fig~\ref{phasediag} for the
limit $S \rightarrow \infty$.
There is a $T=0$ state with long-range spiral spin order for $J_3
> \frac{1}{4} J_1$. We establish that at $0<T<T_c \sim (J_3
- \frac{1}{4} J_1) S^2$ above this state there is a phase with
broken discrete symmetry of lattice reflections about the $x$ and
$y$ axes, while spin rotation invariance is preserved. This phase
has `Ising nematic' order. We present strong numerical evidence
that the transition at $T_c$ is indeed in the Ising universality
class. Such Ising nematic order was originally proposed in
Ref.~\cite{rs} for $S=1/2$ in a $T=0$ spin liquid phase described
by a $Z_2$ gauge theory \cite{z2z2}. Thus the same Ising nematic
order can appear when spiral spin order is destroyed either by
thermal fluctuations (as in the present paper: see
Fig~\ref{phasediag}) or by quantum fluctuations (as in
Ref.~\cite{rs}). Our large $S$ results are therefore consonant
with the possibility of a spin liquid phase at $S=1/2$ as
described in Ref.~\cite{rs,j1j3}; we will discuss the quantum finite
$S$ phase diagram further towards the end of the paper. We also
suggest that discrete lattice symmetries may play a role near
other quantum critical points with spiral order \cite{mnsi}.

Broken discrete symmetries have also been discussed
\cite{ccl,j1j2} in the context of the $J_1{-}J_{2}$ model, with
first- and second-neighbor couplings on the square lattice.
However this model has only collinear, commensurate spin
correlations, and this makes both the classical and quantum theory
quite different from that considered here. As will become clear
below, the spiral order and associated Lifshitz point, play a
central role in the structure of our theory and in the $T$
dependence of observables.

\begin{figure}
\vspace{-0mm}
\includegraphics[width=0.35\textwidth]{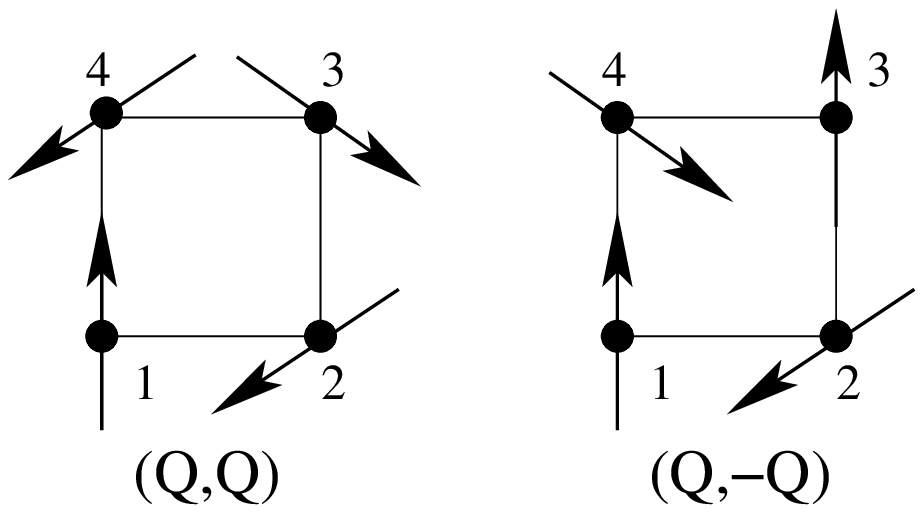}
\vspace{0mm} \caption{\label{fig0} The two different minimum
energy configurations with magnetic wave vectors $\vec{Q}=(Q,Q)$
and $\vec{Q}^\star=(Q,-Q)$ with $Q=2\pi/3$, corresponding to
$J_3/J_1=0.5$. }
\end{figure}

We begin by recalling \cite{locher} the ground states of
$\mathcal{H}$ at $S=\infty$. There is conventional N\'eel order
with magnetic wavevector $\vec{Q}= (\pi,\pi)$ for $J_3/J_1\le
\frac{1}{4}$. For  $J_3/J_1 > \frac{1}{4}$ the ground state has
planar incommensurate antiferromagnetic order, with a pitch vector
depending on the frustration ratio. More precisely, the magnetic
ordering can be described by a wavevector  $\vec{Q}=(Q,Q)$ with
$Q$ decreasing from $\pi$ as $J_3/J_1> \frac{1}{4}$ and
approaching $Q=\pi/2$ monotonically for $J_3/J_1\to \infty$; at
$J_3=\infty$ we obtain four decoupled N\`eel lattices. The spiral
order is in general incommensurate for $\frac{1}{4} < J_3/J_1 <
\infty$, with the exclusion of $J_3/J_1=0.5$ where $Q=2\pi/3$,
corresponding to an angle of $120^\circ$ between spins (See
Fig.~\ref{fig0}). Interestingly, for each spiral state with
$\vec{Q}=(Q,Q)$ there is an energetically equivalent configuration
with $\vec{Q}^\star=(-Q,Q)$, which is distinct from the first one
for $Q\ne \pi$. This state cannot be obtained from the one with
wavevector $\vec{Q}$ by a global rotation of all the spins.
Instead, the two configurations are connected to each other by a
global rotation combined with a reflection about the $x$ or $y$
axes, so that the global symmetry of the classical ground state is
O(3)$\times Z_2$, with an additional two-fold degeneracy beyond
that of the N\`eel case.

\begin{figure}
\vspace{-0mm}
\includegraphics[width=0.41\textwidth]{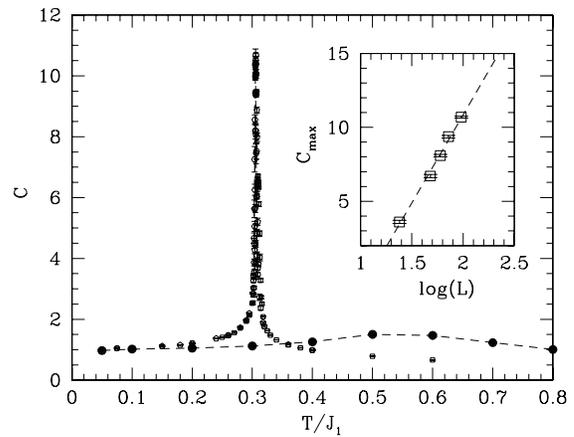}
\vspace{-2mm} \caption{\label{sph} $T$ dependence of the specific
heat for $J_3/J_1=0.5$. Different symbols refer to different
clusters with liner size between $L=24$ and $L=120$. Data for
$J_3/J_1=0.1$ are shown for comparison (full dots and dashed
line). Inset: size-scaling of the maximum of the specific heat. }
\end{figure}

One of the main claims in Fig~\ref{phasediag} is that the broken
$Z_2$ symmetry survives for a finite range of $T>0$, while
continuous O(3) symmetry is immediately restored at any non-zero
$T$, as required by the Mermin-Wagner theorem \cite{mermin}. We
established this by extensive Monte Carlo simulation using a
combination of Metropolis and over-relaxed algorithm for periodic
clusters of size up to $M=120\times 120$, and for several values
of $J_3/J_1$ between 0.25 and 4. Indeed, the presence of a finite
$T$ phase transition is clearly indicated by a sharp peak of the
specific heat which is illustrated in Fig.~\ref{sph}. \cite{note2} This sharp
feature is to be contrasted to the broad maximum displayed by the
same quantity for $J_3/J_1 < \frac{1}{4}$, {\em i.e.\/}, when the
classical ground-state displays ordinary N\`eel order. In
particular, the maximum of the specific heat is consistent with a
logarithmic dependence on system size (see the inset of
Fig.~\ref{sph}) corresponding to a critical exponent $\alpha=0$,
in agreement with Ising universality.
\begin{figure}
\includegraphics[width=0.41\textwidth]{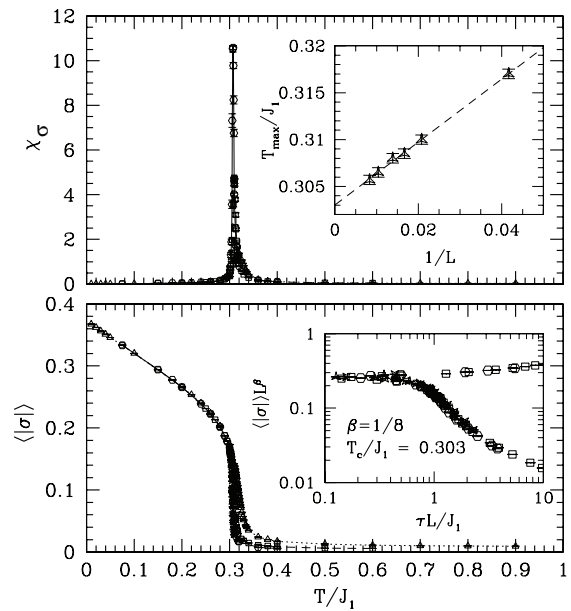}
\vspace{-0mm} \caption{\label{opar} Bottom: $T$ dependence of the
order parameter, $\sigma$, (see Eq.~(\ref{defsigma})) for
different cluster sizes and $J_3/J_1=0.5$. The inset shows the
data collapse according to the scaling hypothesis with Ising
exponents $\beta=1/8$ and $\nu=1$, and $T_c=0.303$. Top:
Temperature dependence of the susceptibility of $\sigma$ for
$J_3/J_1=0.5$. The inset shows the size scaling of the maximum of
the susceptibility. }
\end{figure}

This critical behavior can be directly related to the broken
lattice reflection symmetry by studying an appropriate Ising
nematic order parameter. From the symmetries of Fig~\ref{fig0}, we
deduce that the order parameter is $\sigma = 1/M \left( \sum_a
\sigma_a \right) $ with
\begin{equation}
\sigma_a = \left( \hat{\bf {S}}_{1}\cdot\hat{\bf {S}}_{3}
-\hat{\bf {S}}_{2}\cdot\hat{\bf {S}}_{4} \right)_a~,
\label{defsigma}
\end{equation}
where $a$ labels each plaquette of the square lattice and
$(1,2,3,4)$ are its corners. The variables $\sigma_a$ are
identically zero for a N\`eel antiferromagnet, while they assume
opposite signs on the two degenerate ground states in the spiral
phase. In a phase with all the symmetries of the Hamiltonian,
$\langle \hat{\bf {S}}_{1}\cdot\hat{\bf {S}}_{3} \rangle = \langle
\hat{\bf {S}}_{2}\cdot\hat{\bf {S}}_{4} \rangle $ so that $\langle
\sigma_a \rangle$ vanishes; a phase with Ising nematic order is
signaled by a $\langle \sigma_a \rangle \neq 0$.

Indeed, as clearly shown in Fig.~\ref{opar} (lower panel), the
critical behavior signaled by the divergence of the specific heat
corresponds to a continuous phase transition between a low $T$
phase with a finite value of $\sigma$ in the thermodynamic limit,
and a homogeneous high $T$ phase. Such a transition is also
clearly evidenced by a sharp divergence of the susceptibility of
the Ising nematic order parameter defined by $\chi_\sigma = (M/T)
\left (\langle \sigma^2 \rangle - \langle | \sigma | \rangle^2
\right)$ (see Fig.~\ref{opar}, upper panel), and also by the $T$
dependence of Binder's fourth cumulant $U_4 = 1 - \langle \sigma^4
\rangle / 3 \langle \sigma^2 \rangle^2$ (not shown).

The critical exponent $\nu$ can be easily estimated from the size
dependence of the $T$ corresponding to the maximum of the
susceptibility, which is expected to scale as $T_{max}(L)=T_c + a
L^{-1/\nu}$, where $T_c$ is the thermodynamic critical
temperature. As shown in Fig.~\ref{opar} (upper inset), no sizable
deviation from the Ising exponent $\nu=1$ is observed.  The
exponent $\beta$ is also in agreement with the Ising universality
class. This can be extracted by performing a size scaling analysis
of the order parameter around the critical $T$. In fact, according
to the scaling hypothesis, close to $T_c$, $|\sigma| = L
^{-\beta/\nu}f(x)$, where $f(x)$ is the scaling function, and $x =
\tau L^{1/\nu}$ with $\tau=|T-T_c|$. We have therefore plotted
$\sigma L^{\beta/\nu}$ as a function of $x$, using the value of
the critical temperature $T_c = 0.303(1)$ that can be estimated
from the position of the maximum of the susceptibility
(Fig.~\ref{opar}, upper inset), and the behavior of the Binder's
cumulant. As shown in Fig.~\ref{opar} (lower inset), an excellent
collapse of the data for different lattice sizes on the same curve
is obtained for $\beta/\nu = 1/8$.

\begin{figure}
\vspace{-15mm}
\includegraphics[width=0.41\textwidth]{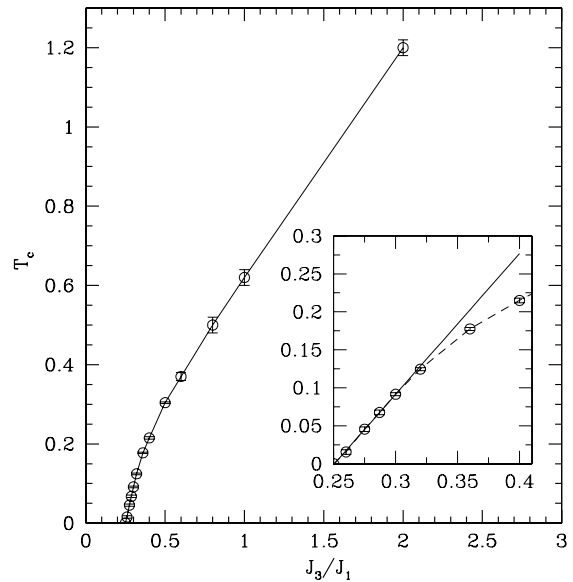}
\vspace{-0mm}
\caption{\label{tcrit}
Critical temperature as a function of the frustration ratio $J_3/J_1$.
}
\end{figure}

We have repeated a similar analysis for several values of
$J_3/J_1$ and the complete phase diagram is shown in
Fig.\ref{tcrit}, where we have plotted $T_c$ {\em vs} $J_3/J_1$.
We find that $T_c$ vanishes linearly for $J_3/J_1\to 1/4$; a
theory for this behavior will now be presented.

Near the classical Lifshitz point, we can model quantum and
thermal fluctuations by a continuum unit vector field ${\bf n} (r,
\tau)$, where $r=(x,y)$ is spatial co-ordinate, $\tau$ is
imaginary time, and ${\bf n}^2 =1 $ at all $r$, $\tau$. This field
is proportional the N\'{e}el order parameter with $\hat{\bf S}_j
\propto (-1)^{x_j + y_j} {\bf n} (r_j, \tau)$. Spiral order will
therefore appear as sinusoidal dependence of ${\bf n}$ on $r$. The
action for ${\bf n}$ is the conventional O(3) non-linear sigma
model, expanded to include quartic gradient terms
($\hbar=k_B=\mbox{lattice spacing}=1$): $\mathcal{S}_{\bf n} =
\int_0^{1/T} d\tau \int d^2 r \mathcal{L}_{\bf n}$ with
\begin{eqnarray}
\mathcal{L}_{\bf n} &=& \frac{\chi_\perp}{2} \left( \partial_\tau
{\bf n} \right)^2 + \frac{\rho}{2} \left[ \left(
\partial_x {\bf n} \right)^2 + \left( \partial_y
{\bf n} \right)^2 \right] \nonumber \\
&+& \frac{\zeta_1}{2} \left[ \left(
\partial_x^2 {\bf n} \right)^2 + \left( \partial_y^2
{\bf n} \right)^2 \right] + \zeta_2
\partial_x^2 {\bf n} \cdot \partial_y^2
{\bf n} \label{ln} \\ &~& \!\!\!\!\!\!\!\!\!\!\!\!\!\!\!\!\!\!\!\!
+\lambda_1 \left[ \left(
\partial_x {\bf n} \cdot \partial_x {\bf n} \right)^2 + \left( \partial_y
{\bf n} \cdot \partial_y {\bf n} \right)^2 \right]  + \lambda_2
\left(
\partial_x {\bf n} \cdot \partial_y {\bf n}
\right)^2 +  \cdots \nonumber
\end{eqnarray}
where the ellipses denote a finite number of additional
$\lambda_i$ couplings involving 4 powers of ${\bf n}$ and 4
spatial derivatives invariant under spin rotations and lattice
symmetries. In the limit $S \rightarrow \infty$ we have
$\chi_\perp = 1/(8J_1)$, $\rho = (J_1 - 4 J_3) S^2$, $\zeta_1 =
(16 J_3 - J_1)S^2/12$, $\zeta_2 =0$, and all $\lambda_i = 0$.
Notice that $\rho$ crosses zero at the Lifshitz point and so can
be regarded as the tuning parameter; $\rho=0$ generally locates
the Lifshitz point for when $\rho < 0$ it is energetically
advantageous to have a $r$-dependent spiral in ${\bf n}$.

A convenient analysis of the properties of $\mathcal{S}_{\bf n}$
is provided by a direct generalization of the $1/N$ expansion of
Ref.~\cite{csy}. The results quoted in Fig~\ref{phasediag} and its
caption were obtained from the $N=\infty$ saddle point equation,
and (apart from certain pre-exponential factors) all functional
forms are exact. The saddle point implements the constraint ${\bf
n}^2 = 1$ and takes the form
\begin{equation}
3 T \sum_{\omega_n} \int \frac{d^2 k}{4 \pi^2} \chi_{\bf n} (k,
\omega_n) = 1\;, \label{constraint}
\end{equation}
where $k$ is a wavevector, $\omega_n$ is a Matsubara frequency,
and $\chi_{\bf n}$ is the dynamic staggered spin susceptibility
with
\begin{eqnarray}
\chi_{\bf n} (k, \omega_n) &=& \Bigl( m^2 + \chi_\perp \omega_n^2
+ \rho (k_x^2 + k_y^2 ) \nonumber \\ && \quad \quad + \zeta_1 (
k_x^4 + k_y^4 ) + 2\zeta_2 k_x^2 k_y^2\Bigr)^{-1} \;.
\label{chival}
\end{eqnarray}
The parameter $m$ is determined by solving Eq.~(\ref{constraint}).

In the classical limit, $S \rightarrow \infty$, we need only
retain the $\omega_n=0$ term in Eq.~(\ref{constraint}) \cite{log}.
A solution for $m$ exists for all $T>0$, and leads to the
crossovers in the spin correlation length $\xi_{\rm spin}$ shown
in Fig~\ref{phasediag}. The value of $\xi_{\rm spin}$, and the
pitch of the spiral order $\sim \sqrt{-\rho}$, as $T \rightarrow
0$ are obtained from the spatial Fourier transform of $\chi_{\bf
n} (k, 0)$.

To investigate the Ising nematic order, we need to study
correlations of the order parameter $\sigma (r, \tau)$ which we
define by a gradient expansion of Eq.~(\ref{defsigma})
\begin{equation}
\sigma = {\bf n} \cdot \partial_x \partial_y {\bf n} - \partial_x
{\bf n} \cdot \partial_y {\bf n}. \label{defsigma2}
\end{equation}
The Ising susceptibility, $\chi_\sigma$ is then $\chi_\sigma =
\int_0^{1/T} d \tau \int d^2 r \langle \sigma (r, \tau) \sigma
(0,0) \rangle$.

In the classical limit, $S\rightarrow \infty$, important exact
properties of $\chi_\sigma$ follow from the ultraviolet finiteness
of the two-dimensional field theory with Boltzmann weight
$\exp\left( - (1/T) \int d^2 r \mathcal{L}_{\bf n} \right)$ and
${\bf n}$ independent of $\tau$. Under a length rescaling analysis
of this theory in which the $\zeta_{i}$ and $\lambda_i$ are fixed,
we see that both $T$ and $\rho$ scale as inverse length squared.
These scaling dimensions establish that in the classical limit
\begin{equation}
\xi_{\rm spin} = \sqrt{\zeta_1/T} \Phi_1
(\rho/T)~~~;~~~\chi_\sigma = \zeta_1^{-1} \Phi_2 ( \rho/T)
\;,\label{scalechi}
\end{equation}
where $\Phi_i$ are cut-off independent scaling function which
depends only on ratios of the $\zeta_i$ and $\lambda_i$. The Ising
phase transition is associated with a divergence of $\Phi_2$ at
some negative argument of order unity, and Eq.~(\ref{scalechi})
then implies the $T_c \sim -\rho$ dependence shown in
Fig~\ref{phasediag}, and verified numerically in Fig~\ref{tcrit}.
We can also compute $\chi_\sigma$ (including the quantum $\omega_n
\neq 0$ modes) in the large $N$ limit and obtain
\begin{equation}
\chi_\sigma = 24 T \sum_{\omega_n} \int \frac{d^2 k}{4 \pi^2}
k_x^2 k_y^2 \chi_{\bf n}^2 (k, \omega_n)\;. \label{chilargeN}
\end{equation}
Using the results in Eq.~(\ref{chival}), Eq.~(\ref{chilargeN})
predicts an exponential divergence in $1/T$ as $T \rightarrow 0$
for $\rho< 0$. This is, of course, an artifact of the large $N$
limit, as our Monte Carlo studies clearly show that $\chi_\sigma$
diverges with a power-law Ising exponent at a $T_c > 0$.

We turn now to a discussion of the quantum physics at finite $S$.
A key feature again emerges from an analysis of
Eqs.~(\ref{constraint}) and (\ref{chival}), while retaining the
full frequency summation: the soft spin spectrum ($ \omega \sim
k^2$) at the Lifshitz point implies that there cannot be
long-range magnetic order over a finite regime of parameters for
all finite $S$ \cite{il}. After evaluating the frequency integral
at $T=0$, a solution with $m$ real exists for a range of values of
$|J_3 - \frac{1}{4} J_1|$ smaller than $\sim e^{- \tilde{c} S}$,
implying there is a spin gap in this regime. We can reasonably
expect that the Ising nematic order survives into at least a
portion this spin gap phase, as it does at $T>0$.

A more careful analysis of the spin gap phase requires
consideration of Berry phases \cite{dcq,rs}, which are absent in
$\mathcal{L}_{\bf n}$. Assuming second order quantum critical
points, with increasing $J_3$, we first expect a spin gap state
with valence bond solid (VBS) order and confined $S=1/2$ spinon
excitations after leaving the collinear N\'eel state. Conversely,
decreasing $J_3$ from the spiral spin ordered phase, we expect a
$Z_2$ spin liquid with Ising nematic order and deconfined bosonic
spinons \cite{rs,z2z2}. So quite remarkably, we expect the
following sequence of 4 phases to appear {\em for all
half-odd-integer spin $S$\/} with increasing $J_3$: N\'{e}el
LRO--VBS--$Z_2$ spin liquid--spiral LRO. The 2 intermediate phases
have a spin gap, and they appear in a window which is
exponentially small in $S$ for large $S$; the latter 2 phases have
Ising nematic order. Theories for the 3 quantum critical points
between these 4 phases appear in Refs.~\cite{dcq,ssrev}. Of
course, we cannot rule out the possibility that the some of these
critical points and intermediate phases are pre-empted by a first
order transition.

It is interesting to note that other $Z_2$ spin liquids with
fermionic $S=1/2$ spinons have been proposed \cite{wen}, in which
the ground state does {\sl not\/} have Ising nematic order. Our
present results naturally suggest a spin gap state with Ising
nematic order, and mean field theories for such states have only
been obtained with bosonic spinons \cite{rs}. Further studies of
Ising nematic order in quantum spin models will therefore be
valuable in resolving the nature of the spin liquid state.

We acknowledge useful discussions with F.~Becca, D.~Scalapino,
S.~Sorella, and S.~White. This work was supported by the NCSA
under Grant No. DMR020027, the NSF under grants DMR-0098226 (S.S.)
and DMR-0210790, PHY-9907949 at the Kavli Institute for
Theoretical Physics, and by the John Simon Guggenheim Memorial
Foundation (S.S.). Kind hospitality provided by INFM-Democritos
and SISSA is gratefully acknowledged (L.C.).


\begin{thebibliography}{99}
\bibitem{z2z2} Note that the global $Z_2$ symmetry of the Ising
nematic order has no direct connection with the $Z_2$ gauge theory
of the spin liquid phase.
\bibitem{note2} In the discussion of the Monte Carlo
results for the $S\to\infty$ limit we rescale the spin lengths
with $S$.
\end{thebibliography}
\end{document}